\documentclass[aps,pra,reprint,groupeaddress,longbibliography]{revtex4-1}

\usepackage[utf8]{inputenc}
\usepackage[english]{babel}
\usepackage[colorlinks=true,linkcolor=blue,citecolor=blue,urlcolor=blue]{hyperref}
\usepackage{amssymb}
\usepackage{amsmath}
\usepackage{amsfonts}
\usepackage{braket}
\usepackage{esdiff}
\usepackage{nicefrac}
\usepackage{bbold}
\usepackage{graphicx}
\usepackage[table]{xcolor}

\begin{document}


\title{Generating entangled quantum microwaves in a Josephson-photonics device}



\author{Simon Dambach}
\affiliation{Institute for Complex Quantum Systems and Center for Integrated Quantum Science and Technology, Ulm University, Albert-Einstein-Allee 11, 89081 Ulm, Germany}
\author{Björn Kubala}
\affiliation{Institute for Complex Quantum Systems and Center for Integrated Quantum Science and Technology, Ulm University, Albert-Einstein-Allee 11, 89081 Ulm, Germany}
\author{Joachim Ankerhold}
\affiliation{Institute for Complex Quantum Systems and Center for Integrated Quantum Science and Technology, Ulm University, Albert-Einstein-Allee 11, 89081 Ulm, Germany}


\date{March 28, 2017}

\begin{abstract}
When connecting a voltage-biased Josephson junction in series to several microwave cavities, a Cooper-pair current across the junction gives rise to a continuous emission of strongly correlated photons into the cavity modes. Tuning the bias voltage to the resonance where a single Cooper pair provides the energy to create an additional photon in each of the cavities, we demonstrate the entangling nature of these creation processes by simple witnesses in terms of experimentally accessible observables. To characterize the entanglement properties of the such created quantum states of light to the fullest possible extent, we then proceed to more elaborate entanglement criteria based on the knowledge of the full density matrix and provide a detailed study of bi- and multipartite entanglement. In particular, we illustrate how due to the relatively simple design of these circuits changes of experimental parameters allow one to access a wide variety of entangled states differing, e.g., in the number of entangled parties or the dimension of state space. Such devices, besides their promising potential to act as a highly versatile source of entangled quantum microwaves, may thus represent an excellent natural testbed for classification and quantification schemes developed in quantum information theory.
\end{abstract}


\maketitle

\section{Introduction}
\label{sec_Introduction}

The concept of entanglement is at the heart of quantum physics: as one of the cornerstones at the foundation of quantum mechanics and, at the same time, as a key ingredient in emerging quantum technologies introducing a new quantum resource into communication, computing, and sensing.

These two interconnected aspects of entanglement are reflected in a branching of research interests. On one hand, a more abstract, mathematical direction of quantum information theory maps out the boundary between classical and quantum world and further charts quantum states into various classes of entanglement. More and more sophisticated (qualitative) witnesses and (quantitative) measures of entanglement are devised in this field \cite{Bruss2002,Terhal2002,Werner2002,Braunstein2005,Guehne2009,Horodecki2009,Mintert2005,Plenio2007} with many open questions remaining, in particular, concerning mixed states and multipartite systems~\cite{Duer2000,Acin2001}. Oftentimes research considers complex quantum states characterized by certain, special classes of density matrices, for which entanglement witnesses or measures can be worked out, while typically the question in which actual systems and under which actual circumstances such states may naturally be encountered garners less attention.

On the other hand, proposed and realized entanglement sources in the optical and microwave regime typically strive to realize particularly simple entangled states, for instance, pure (squeezed) Gaussian states with small, negligible admixtures~\cite{Villar2005,Flurin2012,Menzel2012}, NOON states~\cite{Afek2010,Wang2011,Su2014}, or maximally entangled states~\cite{Rauschenbeutel2000,Lu2007,Eibl2004}, thereby accessing only a small number of all interesting types of entangled states. In the microwave regime, besides setups using analogues of typical nonlinear optical elements~\cite{Menzel2012}, more recent schemes extend the ability to create arbitrary quantum states in a microwave resonator~\cite{Hofheinz2009} to several, spatially separated modes~\cite{Wang2011}. While by properly choosing an elaborate pulse scheme, in principle, any entangled state could be created, it is typically a maximally entangled or another simple pure entangled state which such experiments aim for.

The creation of multipartite or other more complex entangled states requires increasingly complex pulse schemes. These are reachable in such systems since one can build on the immense research effort (and the resulting amazing progress in performance and control) which has been spent on these standard circuit-quantum-electrodynamics (QED) setups as part of the larger quest for universal quantum computing. Here we argue, however, that combining two key elements of circuit-QED setups, namely, Josephson junctions and microwave cavities, in a much simpler, less demanding device can offer an alternative, fully tunable and versatile entanglement-generating source.

Recently developed Josephson-photonics devices~\cite{Hofheinz2011,Chen2011,Dykman2012,Chen2014}, which already demonstrated their potential as a source of nonclassical microwave light~\cite{Grimm2015,Parlavecchio2015,Parlavecchio2016,Gramich2013,Armour2013,Leppaekangas2013,Kubala2015,Armour2015,Trif2015,Dambach2015,Dambach2016,Kubala2016,Leppaekangas2015,Leppaekangas2016,Souquet2016}, seem to be logical candidates for this task. In fact, the next generation of setups currently being under fabrication is designed
 to explore bipartite photon creation processes. The goals of the present work are thus twofold: to make detailed and quantitative predictions about entanglement properties of this new class of superconducting devices and to give instructions how to vary and design experimental parameters accordingly. In this respect, the simple bipartite case is only the first step. As we will show, Josephson-photonics architectures allow one by comparatively simple changes of experimental parameters to also access multipartite situations, the characterization of which requires the most sophisticated classification and quantification schemes developed in abstract quantum information theory.

The nontrivial entangled states, which appear as (quasi-)stationary states of these voltage-driven, damped systems, allow one to explore the richness of entanglement phenomena along several dimensions: the number of entangled parties, continuous or discrete degrees of freedom, fully entangled or biseparable states, all this is controllable in a single experimental setup.
Josephson-photonics devices or other entanglement sources utilizing similar entanglement creation mechanisms, hence, highlight the necessity of better bridging the gap between the abstract quantum-information branch of research on entanglement and more practical-oriented approaches aimed at utilizing entanglement as a resource for various quantum technological applications.

In this paper, we will first briefly discuss a basic example of realizing either bi- or tripartite entanglement by a simple change of the voltage bias applied to the Josephson junction (Sec.~\ref{sec_Model}). After having thus demonstrated that such a device can, in principle, be used as a source of genuinely multipartite entangled microwaves, we will explain how also other characteristics of entanglement can be explored by varying other experimental ``knobs''. A detailed discussion of bi- and multipartite entanglement in Secs.~\ref{sec_Bipartite_entanglement} and \ref{sec_Multipartite_entanglement}, respectively, then reveals the full richness of entanglement phenomena in Josephson-photonics devices. Limitations which occur in current experimental setups and possible improvements and modifications which might thus become necessary are discussed in Sec.~\ref{sec_Experimental_situation}. We conclude in Sec.~\ref{sec_Conclusions_and}  and address open questions which remain for future research.

\begin{figure*}
\centering
\includegraphics[width=1.0\textwidth]{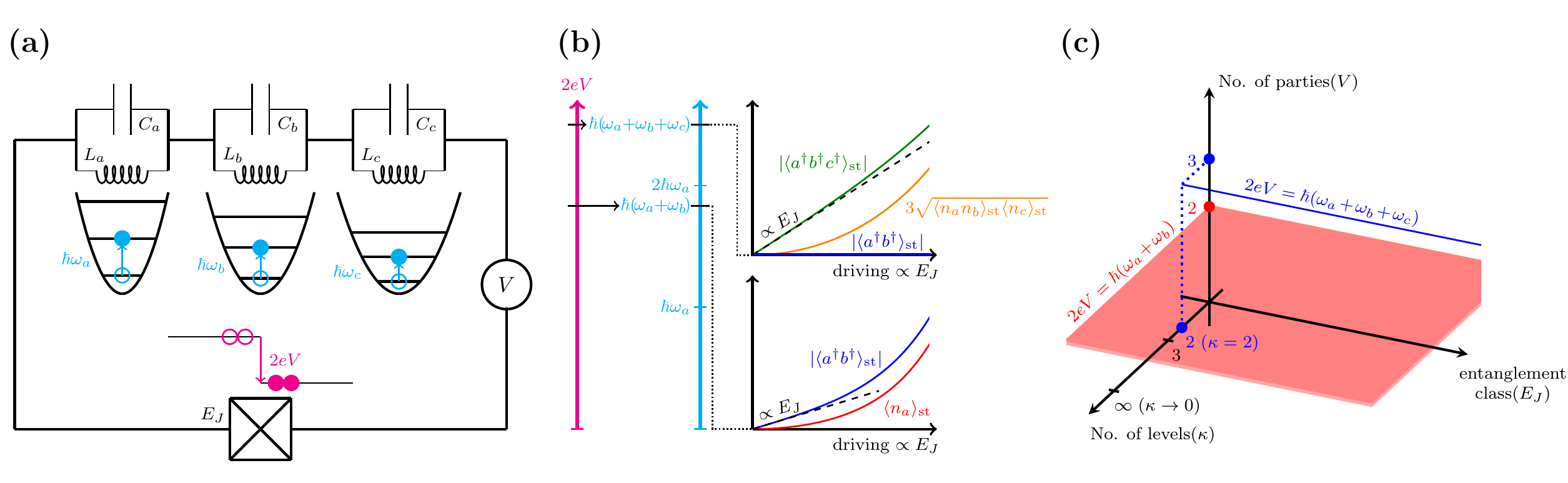}
\vspace{-0.45cm}
\caption{
(a)~Sketch of an effective circuit model consisting of a voltage-biased Josephson junction connected in series to three LC oscillators with frequencies $\omega_{q}=1/\sqrt{L_{q}C_{q}}$. By tuning the external voltage, we can access a variety of different resonances where a single Cooper pair tunneling across the junction provides the energy to excite one or several photons in one, two, or three resonators.
(b)~We focus here on the biasing conditions $\omega_{J}=2eV/\hbar=\omega_{a}+\omega_{b}(+\omega_{c})$, where two (three) photons are simultaneously created in different oscillators. Simple entanglement witnesses introduced in Eq.~\eqref{eq_witness_bipartite} [Eq.~\eqref{eq_witness_tripartite}] prove the presence of genuine bipartite (tripartite) entanglement between the two (three) oscillator subsystems for a wide range of driving.
(c)~Cartoon of the space of entanglement phenomena illustrating the versatile entangling properties of the Josephson-photonics device. Each of the directions (number of parties, number of levels, entanglement class) is associated with an experimentally accessible parameter (voltage $V$, $\kappa$ parameter, Josephson energy $E_{J}$). The nature of entanglement in the red and blue regions is analyzed in detail in Sec.~\ref{sec_Bipartite_entanglement} (Fig.~\ref{fig_Fig_2}) and Sec.~\ref{sec_Multipartite_entanglement} (Fig.~\ref{fig_Fig_5}), respectively.
}
\label{fig_Fig_1}
\end{figure*}

\section{Josephson-photonics device as entanglement source}
\label{sec_Model}

A Josephson-photonics setup, as pioneered by the experimental groups at Saclay/Grenoble~\cite{Hofheinz2011,Grimm2015,Parlavecchio2015} and Dartmouth~\cite{Chen2011,Chen2014,Dykman2012} and subsequently extensively investigated theoretically \cite{Dykman2012,Armour2013,Gramich2013,Armour2015,Trif2015,Kubala2015,Dambach2015,Dambach2016,Kubala2016,Leppaekangas2013,Leppaekangas2015,Leppaekangas2016,Souquet2016, Padurariu2012,Mendes2015}, uses a Josephson junction biased by an external dc voltage $V$
to create microwave excitations in two or more series-connected LC oscillators with
frequencies $\omega_{q}=1/\sqrt{L_{q}C_{q}}$, see Fig.~\hyperref[fig_Fig_1]{1(a)}. These oscillators parallel the microwave stripline cavities coupled to qubits in standard circuit-QED setups, where, however, there is no dc-current path through the system.

Biasing the junction so that its Josephson frequency $\omega_{J}=2eV/\hbar$ matches the sum of the modes frequencies, $\omega_J=\sum_{q}\omega_{q}$, the transfer of a single Cooper pair  across the Josephson junction  gives rise to a simultaneous creation (or annihilation) of one photon within each of the modes.
This common creation process is balanced by subsequent individual leakage of photons via output lines connected to each of the different resonators so that eventually a stationary~\footnote{Besides the leakage of photons from the resonator, local voltage fluctuations at the junction are a second source of decoherence in an actual experimental realization as will be discussed below in Sec.~\ref{sec_Experimental_situation}. The stationary states resulting from the current simplified analysis will appear as quasistationary states when weak voltage fluctuations are included. As discussed in Sec.~\ref{sec_Experimental_situation}, these states are experimentally accessible on intermediate timescales.} (and possibly entangled) state of the spatially separated cavities is reached.

Formally, the particular resonant processes at the given bias are picked out by a rotating-wave approximation after the system Hamiltonian is transformed to a frame rotating with the Josephson frequency $\omega_{J}$~\cite{Armour2013,Gramich2013,Armour2015,Trif2015}. The resulting effective time-independent Hamiltonian
\begin{equation}
H=\frac{E^{*}_{J}}{2}:\Big(\prod_{q}a^{\dagger}_{q}+\prod_{q}a_{q}\Big)\prod_{q}\frac{J_{1}\left(\sqrt{4\kappa_{q}n_{q}}\right)}{\sqrt{\kappa_{q}n_{q}}}:
\label{eq_Hamiltonian_RWA}
\end{equation}
contains a renormalized Josephson energy, $E^{*}_{J}=E_{J}\prod_{q}\sqrt{\kappa_{q}}e^{-\kappa_{q}/2}$, and normal-ordered Bessel functions $J_{1}$ of the first kind. These are functions of the photonic number operators $n_{q}=a_{q}^{\dagger}a_{q}$ of the various resonators, expressed in terms of creation/annihilation operators with $[a_{q},a_{q}^{\dagger}]=1$, and the dimensionless parameter $\kappa_{q}=(2e^{2}/\hbar)\sqrt{L_{q}/C_{q}}$  measuring each oscillator's zero-point quantum fluctuations. In the Appendix~\hyperref[Appendix_A]{A}, we discuss the equivalent structure of the Hamiltonian for other resonances, e.g., if the chosen bias allows single photon excitations in some subset of all coupled oscillators.

The bracketed term in Eq.~\eqref{eq_Hamiltonian_RWA} indicates the fundamental photon creation/absorption process considered here: each forward (backward) tunneling event of a Cooper pair goes along with the simultaneous creation $\prod_{q}a_{q}^{\dagger}$ (absorption $\prod_{q}a_{q}$) of one photon in each of the oscillator modes.
These fundamental processes, however, are modified by the inherent nonlinearity of the Josephson junction, reflected in the appearance of the product of Bessel functions, which makes the driving nonlinear.

Generally speaking, the effect of these nonlinearities on the dynamics of the system is governed by two different experimental parameters, the $\kappa_{q}$ parameter(s) and the Josephson energy $E_{J}$. The former fixes the relative size of the transition matrix elements between neighboring states of oscillator $q$ and thus effectively sets its level structure as we will see below in Sec.~\ref{subsec_Restricted_Hilbert}. The latter determines the driving strength and thus the population of the corresponding levels.  As we will see later on [cf. Fig.~\hyperref[fig_Fig_1]{1(c)}], varying these parameters (and the resonance chosen) makes it possible to explore a wide range of entanglement phenomena in Josephson-photonics devices.

Creating and annihilating photons according to the Hamiltonian in Eq.~\eqref{eq_Hamiltonian_RWA}, a stationary state is reached in due course as excited photons leak out of the resonators into the electromagnetic environment after a lifetime $1/\gamma_{q}$. Focusing on the zero-temperature limit, this dynamics can be described by a quantum master equation of the standard Lindblad form~\cite{Breuer2002}
\begin{equation}
\diff{\rho}{\tau}\!=\!\mathfrak{L}\rho\!=\!-\frac{i}{\hbar}\left[H,\rho\right]+\sum_{q}\frac{\gamma_{q}}{2}\left(2a_{q}\rho a_{q}^{\dagger}\!-\!n_{q}\rho\!-\!\rho n_{q}\right)
\label{eq_quantum_master_equation}
\end{equation}
for the density operator $\rho$ of the cavity degrees of freedom.

A first basic example demonstrating the entanglement power of the device and at the same time illustrating the general strategy pursued in our investigation is sketched in Fig.~\hyperref[fig_Fig_1]{1(b)}. Let us consider a setup consisting of three cavities $a$, $b$, and $c$, with the photonic operators now denoted $a^{\dagger}_{1}=a^{\dagger}$, $a^{\dagger}_{2}=b^{\dagger}$, and $a^{\dagger}_{3}=c^{\dagger}$. Selecting two biasing conditions $\omega_{J}=2eV/\hbar=\omega_a+\omega_b\, (+ \omega_c)$ and consequently two different resonant excitation processes, the driving is turned up to investigate if (and what type of) entanglement is achieved.

What statements on entanglement can be made depends on the choice of entanglement witness or measure. Later on, we will present a number of more elaborate witnesses designed to optimize the information on entanglement gained. For the moment, however, we consider a particularly simple witness experimentally accessible via measurements on the output lines. Namely, the violation of the inequalities~\cite{Hillery2006,Woelk2014}
\begin{align}
| \langle a^\dagger b^\dagger \rangle_\mathrm{st} | &\le \sqrt{\langle n_a \rangle_\mathrm{st} \langle n_b \rangle_\mathrm{st}}\stackrel{\mathrm{sym.}}{=} \langle n_a \rangle_\mathrm{st}\label{eq_witness_bipartite}\\
\begin{split}
| \langle a^\dagger b^\dagger c^\dagger \rangle_\mathrm{st} | &\le \sqrt{\langle n_a n_b \rangle_\mathrm{st} \langle n_c \rangle_\mathrm{st}}  + \mathrm{perm.}\\&\stackrel{\mathrm{sym.}}{=} 3 \sqrt{\langle n_a n_b \rangle_\mathrm{st} \langle n_c \rangle_\mathrm{st}}\label{eq_witness_tripartite}
\end{split}
\end{align}
bears witness to the presence of genuine bi- and tripartite entanglement for the full range of driving shown in Fig.~\hyperref[fig_Fig_1]{1(b)}.
The equality signs hold for the symmetric case, $\kappa_a = \kappa_b \,(=\kappa_c)$ and $\gamma_a = \gamma_b \,(=\gamma_c)$, also assumed in the figure.

Indeed, we show data for small $\kappa_{a/b/c}=0.1$ so that nonlinearities appear for large occupation numbers only. In the weak driving case, there are but small corrections to the bare entangling process of a nondegenerate parametric amplifier with driving $\propto E_J a^\dagger b^\dagger + \mathrm{c.\,c}$ and its three-party equivalent. In consequence, for the $\omega_{J}=\omega_a + \omega_b$ resonance we immediately recognize in Fig.~\hyperref[fig_Fig_1]{1(b)} the known parametric-amplifier results, $ \langle n_a \rangle_\mathrm{st} = \langle n_b \rangle_\mathrm{st} \propto E_J^2$ and $\langle a^\dagger b^\dagger \rangle_\mathrm{st} \propto E_J$ for weak driving~\cite{Armour2015,Dambach2016}. If the inequality holds, no statement on entanglement is possible based on the chosen witness.
Similarly, for the $\omega_{J}=\omega_a + \omega_b+\omega_c$ resonance we cannot witness entanglement between two modes alone ($\langle a^\dagger b^\dagger \rangle_\mathrm{st}\equiv0$), but instead all three resonators are entangled for the whole driving range shown.

These simple cases can now be used to exemplify the key strategy of this investigation. The versatile entangling properties of Josephson-photonics devices are exploited  to explore the space of entanglement phenomena along several directions. Schematically, this is visualized in Fig.~\hyperref[fig_Fig_1]{1(c)}. Roughly speaking, different experimental knobs in our setup correspond to different dimensions of the complete space of all possible entangled states:\\
(i)~Changing the bias voltage, the number of actively involved oscillators and thereby the number of parties which are (potentially) entangled is varied.\\
(ii)~Changing the parameter $\kappa_q$, different effective Hilbert (sub-)spaces for party $q$ are realized. For $\kappa_q \rightarrow 0$, for instance, all levels of oscillator $q$ are accessible, while for special values of $\kappa_q$ oscillator $q$ reduces to an effective $N$-level state (most importantly $\kappa_{q}=2$ yields a two-level system as discussed in Sec.~\ref{subsec_Restricted_Hilbert} below). \\
(iii)~The driving strength $E_J$ finally is changed to tune the system through different classes of entanglement. These can be as simple as in the basic example above, where our witness only allows the identification of an entangled region below some $E_J$ and an unchartered region above. The aim, however, (achieved in this paper for the cases marked by the blue and red regions in the figure) is the complete characterization of entanglement for the chosen scenario.

As a bottom line, we have demonstrated that the new experimental platform of Josephson-photonics devices has indeed the potential to create entangled photon states. The type of entanglement is determined by a small set of experimentally tunable parameters. To show this in detail, we now turn toward the simplest scenario, namely, bipartite entanglement.

\begin{figure*}
\centering
\includegraphics[width=1.0\textwidth]{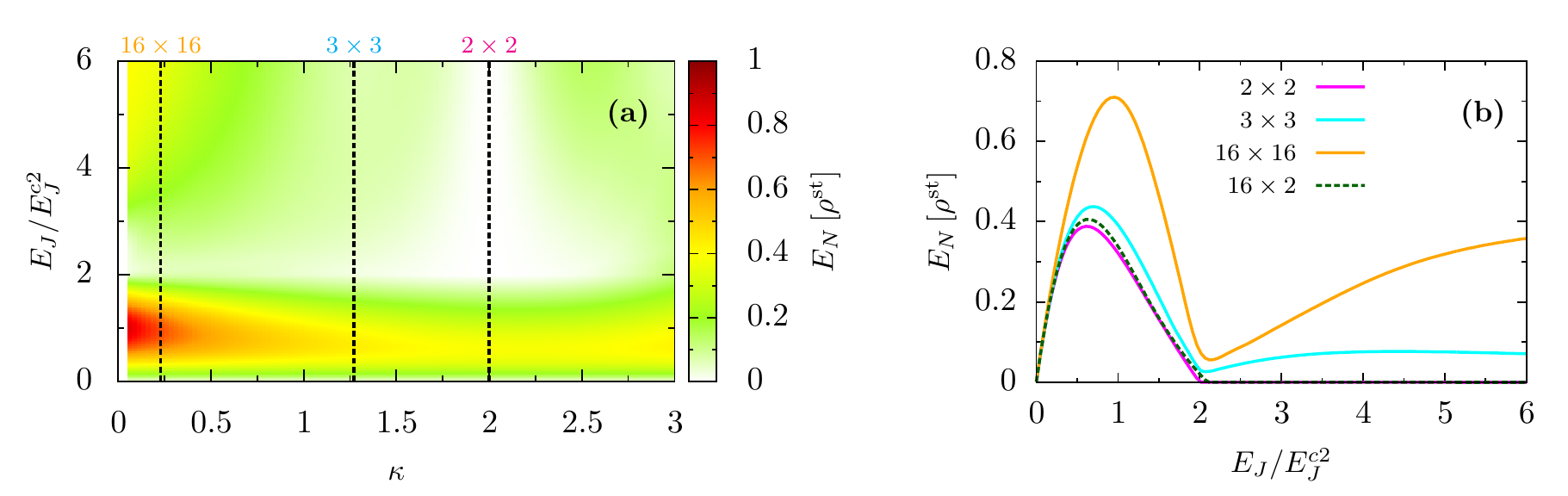}
\vspace{-0.45cm}
\caption{
(a)~Logarithmic negativity $E_{N}$ in steady state for two symmetric oscillators ($\kappa_{a}=\kappa_{b}=\kappa$, $\gamma_{a}=\gamma_{b}=\gamma$) as a function of the $\kappa$ parameter and the driving $E_{J}$. Dashed lines refer to special values of $\kappa$ where the state space of each of the two oscillators is reduced to $2$, $3$, or $16$ levels. (b)~Cross sections of the dashed lines in (a) supplemented by the corresponding results for the $16\times2$ system. Steady-state entanglement vanishes only in an $N\times2$ Hilbert space for sufficiently strong driving.
}
\label{fig_Fig_2}
\end{figure*}

\section{Bipartite entanglement}
\label{sec_Bipartite_entanglement}

In the three-oscillator setup presented above, a bipartite system is realized by selecting a bias condition $2eV/\hbar=\omega_{a}+\omega_{b}$. The corresponding effective Hamiltonian differs from an experimentally already realized two-cavity setup merely in a renormalization of the Josephson energy, see Eq.~\eqref{eq_Hamiltonian_nonactive_cavities} in the Appendix~\hyperref[Appendix_A]{A}.
In any case, the simultaneous creation of a single photon in each of the oscillators $a$ and $b$ leads to strongly correlated dynamics including potentially, but not necessarily, quantum entanglement. These two subsystems are called entangled if their corresponding mixed state $\rho$ is \emph{not} separable, i.e., if it can \emph{not} be written as a convex combination of product states of the subsystems: $\rho=\sum_{j}p_{j}\rho^{j}_{a}\otimes\rho^{j}_{b}$ with convex weights $p_{j}>0$ and $\sum_{j}p_{j}=1$.

The question whether a given bipartite density matrix is separable or entangled is still lacking a general answer. Over the last decades, however, a large number of different criteria have been proposed to detect the presence of entanglement or to prove separability (for an overview see, e.g., Refs.~\cite{Bruss2002,Terhal2002,Werner2002,Braunstein2005,Guehne2009,Horodecki2009}). In experiments, the analysis of entanglement is often restricted to simple witnesses based on directly measurable quantities [see, e.g., Eq.~\eqref{eq_witness_bipartite} in the previous section]. Our theoretical approach, however, provides full information on the system, i.e., the complete density matrix, which is experimentally accessible only by full state tomography, so that in essence the whole range of theoretically established entanglement criteria may be applied.

\subsection{PPT criterion and logarithmic negativity}
\label{subsec_PPT_criterion_and_logarithmic_negativity}
 
An entanglement criterion both powerful and simple is the criterion based on positive partial transposition (PPT criterion)~\cite{Peres1996}. It relies on the fact that taking the transposition with respect to a single subsystem only does not necessarily map a state $\rho$ onto a quantum state again. If, however, $\rho$ is a separable state, its partial transpose $\rho^{T_{a}}=\sum_{j}(\rho^{j}_{a})^{T}\otimes\rho^{j}_{b}$ (and analogous for subsystem $b$) indeed represents a valid density matrix and is thus positive semidefinite, $\rho^{T_{a}}\ge0\Leftrightarrow\rho^{T_{b}}\ge0$. In consequence, we can conclude that $\rho$ is entangled if its partial transpose is not positive semidefinite, i.e., if it has at least one negative eigenvalue. In this case, we say that $\rho$ has a negative partial transpose (NPT). Importantly, the converse is explicitly \emph{not} true in general: a PPT, i.e., the partial transpose of $\rho$ is positive semidefinite, does only imply separability for the special cases of the low-dimensional $3\times2$ and $2\times2$ systems~\cite{Horodecki1996,Horodecki1997}.

To additionally quantify the amount of entanglement of $\rho$, the logarithmic negativity $E_{N}[\rho]=\log_{2}(||\rho^{T_{a}}||_{1})$~\cite{Zyczkowski1998,Vidal2002}, which is directly linked to violation of the PPT criterion, is used. This entanglement measure is based on the trace norm $||\dots||_{1}$ of the partial transpose $\rho^{T_{a}}$, which is related to the sum of the negative eigenvalues $\lambda^{j}_{-}$ of $\rho^{T_{a}}$: $||\rho^{T_{a}}||_{1}=1+2|\sum_{j}\lambda^{j}_{-}|$. PPT criterion as well as logarithmic negativity are easy to calculate for arbitrary-dimensional state spaces if the density matrix is known; however, they obviously fail in detecting some entangled states, namely, those with a positive partial transpose.

We can now proceed to pursue our general strategy and explore the remaining directions of the space of entanglement phenomena (sketched in Fig.~\hyperref[fig_Fig_1]{1(c)} as a plane) for this two-party case of Josephson photonics.
For a first overview, the symmetric $\kappa_q$ direction, i.e., $\kappa_{a}=\kappa_{b}=\kappa$, is chosen and also the damping of the two oscillators is assumed to be equal, $\gamma_{a}=\gamma_{b}=\gamma$. Figure~\hyperref[fig_Fig_2]{2(a)} then shows the driving dependence of logarithmic negativity $E_{N}$, while $\kappa$ is allowed to increase from the parametric-amplifier limit $\kappa\rightarrow 0$ with its harmonic level structure and driving. The results for the entanglement witness shown in Fig.~\hyperref[fig_Fig_1]{1(b)} are close to this limit with $\kappa=0.1$.

Two important features of Fig.~\hyperref[fig_Fig_2]{2(a)} will be explored  in the following discussion. Firstly, the logarithmic negativity $E_{N}$ shows a maximum which becomes the more pronounced the smaller the value of $\kappa$ is with its position monotonically shifting from $E_{J}<E^{c2}_{J}=(\hbar\sqrt{\gamma_{a}\gamma_{b}}/\sqrt{\kappa_{a}\kappa_{b}})e^{(\kappa_{a}+\kappa_{b}+\kappa_{c})/2}$ toward $E_{J}/E^{c2}_{J}\approx1$.
($E^{c2}_{J}$ is the amplification threshold in the parametric-amplifier limit $\kappa\rightarrow 0$.)
Secondly, $E_{N}$ has a distinct minimum at $E_{J}/E^{c2}_{J}\approx2$, nearly independent of the actual value of $\kappa$. At $\kappa=2$, however, $E_{N}$ does not only take a minimum but vanishes and stays zero for all $2\le E_{J}/E^{c2}_{J}$. This feature, specific to $\kappa=2$, is related to the restricted Hilbert space of each of the entangled parties, as the suppressions of any transitions to higher excited levels  effectively reduces the harmonic oscillator to a two-level system.

\subsection{Restricted Hilbert space}
\label{subsec_Restricted_Hilbert}

The special values of $\kappa$ where the state space of each of the two oscillators is effectively reduced to a sixteen- ($\kappa\approx0.23$), three- ($\kappa=3-\sqrt{3}$), or two-level system ($\kappa=2$) are indicated by dashed lines in Fig.~\hyperref[fig_Fig_2]{2(a)} [corresponding to the cross sections in Fig.~\hyperref[fig_Fig_2]{2(b)}].
These special values simply follow from the roots of the transition matrix elements $T_{m_{a},m_{b};m_{a}+1,m_{b}+1}=\bra{m_{a},m_{b}}H\ket{m_{a}+1,m_{b}+1}$ stemming from the (normal-ordered) Bessel functions which captures the inherent nonlinearity of the Josephson junction. The transition matrix element factorizes into matrix elements involving a single cavity only and can be given in terms of the generalized Laguerre polynomials $L^{(1)}_{m_{a/b}}(\kappa)$~\cite{Dambach2015,Souquet2016}.

While results plotted in Fig.~\ref{fig_Fig_2} are found by solving $\mathfrak{L}\rho=0$ numerically, for the low-dimensional case of a $2\times2$ Hilbert space obtaining results in a simple analytical form can provide some interesting insights. Solving the coupled equations of motion for the density matrix elements in the standard product basis $\lbrace\ket{00},\ket{01},\ket{10},\ket{11}\rbrace$, the steady-state density matrix takes the simple form
\begin{equation}
\rho^{\mathrm{st}}=
\begin{pmatrix}
\rho^{\mathrm{st}}_{00,00}&0&0&\rho^{\mathrm{st}}_{00,11}\\
0&\rho^{\mathrm{st}}_{01,01}&0&0\\
0&0&\rho^{\mathrm{st}}_{10,10}&0\\
\rho^{\mathrm{st}}_{11,00}&0&0&\rho^{\mathrm{st}}_{11,11}\\
\end{pmatrix}\!,
\end{equation}
where the nonvanishing matrix elements are a function of $E_{J}$ and $\gamma$.
They are given in the Appendix~\hyperref[Appendix_B]{B} [Eq.~\eqref{eq_analytical_results_two_qubits}] for the general case of asymmetric damping. For equal damping, the logarithmic negativity is only a function of the absolute value of the coherence $\rho^{\mathrm{st}}_{00,11}\equiv(\rho^{\mathrm{st}}_{11,00})^{*}$ and the population $\rho^{\mathrm{st}}_{01,01}\equiv\rho^{\mathrm{st}}_{10,10}$: $E_{N}[\rho^{\mathrm{st}}]=\mathrm{max}\lbrace0,\log_{2}[1+2(|\rho^{\mathrm{st}}_{00,11}|-\rho^{\mathrm{st}}_{01,01})\rbrace$.

The results for $E_{N}$ in Fig.~\ref{fig_Fig_2} are easy to understand for weak driving, $E_{J}/E^{c2}_{J}\rightarrow0$. In that limit,  populations as $\rho^{\mathrm{st}}_{01,01}\propto E_{J}^2$ increase slower than  coherences $|\rho^{\mathrm{st}}_{00,11}|\propto E_{J}$ so that $E_{N}[\rho^{\mathrm{st}}]$ steadily increases with the driving.
For larger values of $E_{J}/E^{c2}_{J}$, however, $\rho^{\mathrm{st}}_{01,01}$ saturates while $|\rho^{\mathrm{st}}_{00,11}|\propto E_{J}^{-1}$ decreases,  hence $E_{N}[\rho^{\mathrm{st}}]$ decreases and eventually vanishes when $|\rho^{\mathrm{st}}_{00,11}|\le\rho^{\mathrm{st}}_{01,01}$ with  $E_{N}[\rho^{\mathrm{st}}]=0$ if $2\le E_{J}/E^{c2}_{J}$. A maximal entanglement of
$E_{N}[\rho^{\mathrm{st}}]=\log_{2}[(3+\sqrt{5})/4]$ is found in between for $E_{J}/E^{c2}_{J}=2/(1+\sqrt{5})$ (the inverse of the golden ratio).

The entanglement features of the $2\times2$ system are passed on  nearly unchanged to systems with unequal $\kappa_q$ chosen such that one party still is a two-level system, while the dimension of the Hilbert space of the second party is increased, e.g., the $16\times2$ realization in Fig.~\hyperref[fig_Fig_2]{2(b)}.
The similarity to the  $2\times2$ case can be explained by the fact that, in consequence of energy conservation and the simultaneous creation of photons, the mean occupations of the two oscillators are coupled, $\gamma_{a}\langle n_{a}\rangle_{\mathrm{st}}=\gamma_{b}\langle n_{b}\rangle_{\mathrm{st}}$. The presence of the qubit thus leads to signatures of state-space restriction in the sixteen-level system (cf. Refs.~\cite{Dambach2016,Souquet2016}), which are reduced ($\gamma_{a}<\gamma_{b}$) or enhanced ($\gamma_{a}>\gamma_{b}$) for asymmetric decay rates.

Note that $E_{N}[\rho^{\mathrm{st}}]=0$ does, in general, \emph{not} imply separability for a $16\times2$ system. However, we know that if an $N\times2$ state $\rho$ is invariant under partial transposition with respect to the qubit, i.e., $\rho^{T_{b}}=\rho$, then $\rho$ is separable~\cite{Kraus2000}. Since the property of separability is independent of invertible local transformations $U$ acting on $\mathbb{C}^{2}$, also $[(\mathbb{1}\otimes U)\rho(\mathbb{1}\otimes U)^{\dagger}]^{T_{b}}=(\mathbb{1}\otimes U)\rho(\mathbb{1}\otimes U)^{\dagger}$ implies separability. Due to the simple structure of $\rho^{\mathrm{st}}$ in the $16\times2$ case where all coherences are zero except for $\rho^{\mathrm{st}}_{j0,(j+1)1}$ with $j\in\lbrace0,\dots,14\rbrace$, we can decompose $\rho^{\mathrm{st}}$ into different legitimate quantum states and find appropriate transformations $U$.  The steady state is then expressed as $\rho^{\mathrm{st}}=(\rho_{+}+\rho_{-})/2$ where $\rho_{\pm}=\rho_{\mathrm{diag}}+\rho_{\mathrm{off}}\pm\rho_{\mathrm{off}}^{T_{b}}$ with $\rho_{\mathrm{diag}}$ and $\rho_{\mathrm{off}}$ being the diagonal and off-diagonal part of $\rho^{\mathrm{st}}$, respectively. Choosing $U_{+}=\mathbb{1}$ ($U_{-}=\mathrm{diag}[1+i,1-i]/\sqrt{2}$) shows that $\rho_{+}$ ($\rho_{-}$) and thus $\rho^{\mathrm{st}}$ is separable. So, indeed, the $16\times2$ system here is separable whenever $E_{N}[\rho^{\mathrm{st}}]=0$.

Separability for strong driving in the $N\times2$ cases does not mean that quantum correlations between the two subsystems are absent. This is reflected in a nonzero quantum discord~\cite{Ollivier2001,Henderson2001}, which measures quantum correlations by the difference between quantum mutual information and classical correlations. States of the $2\times2$ and $16\times2$ system with $E_{N}[\rho^{\mathrm{st}}]=0$ have a strictly positive quantum discord, i.e., they contain nonclassical correlations even though they are separable. This is shown using a simple witness~\cite{Ferraro2010} which allows the detection of nonvanishing quantum discord: If a bipartite state $\rho$ does \emph{not} commute with $\rho_{a}\otimes\mathbb{1}$, where $\rho_{a}=\mathrm{Tr}_{b}\lbrace\rho\rbrace$, then its quantum discord is nonzero and $\rho$ is nonclassically correlated. It has been shown that even such nonclassically correlated states without entanglement may represent a valuable resource for a number of different quantum information processing tasks~\cite{Meyer2000,Datta2007,Lanyon2008}.   

\begin{figure}[b]
\centering
\includegraphics[width=1.0\columnwidth]{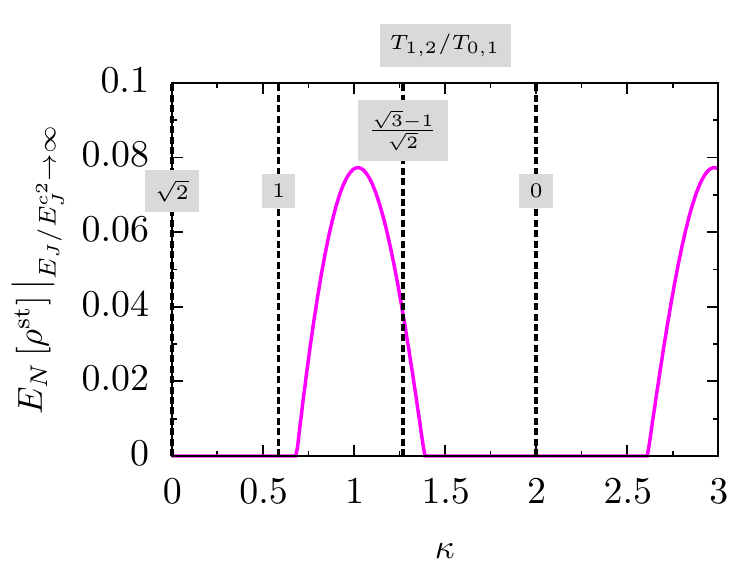}
\vspace{-0.45cm}
\caption{
Steady-state logarithmic negativity $E_{N}$ in the strong-driving limit, $E_{J}/E^{c2}_{J}\rightarrow\infty$. State space is cut to a $3\times3$ system of two symmetric oscillators.
Whether steady-state entanglement is observable  depends on the $\kappa$ parameter, which determines the ratio $T_{1,2}/T_{0,1}$ of the two transition matrix elements. Dashed lines indicate special realizations of $3\times3$ systems (see main text for details).
}
\label{fig_Fig_3}
\end{figure}

\subsection{Modified driving Hamiltonian}

Steady-state entanglement vanishes for strong driving only in case that a qubit is involved. For all other realizations, e.g., the $3\times3$ or $16\times16$ realization, the logarithmic negativity takes a minimum around $E_{J}/E^{c2}_{J}=2$ followed by an increase in $E_{N}$. This increase can be traced back to coherences associated with higher photon creation processes, e.g., $\rho_{00,22}$ in the $3\times3$ case. These do not vanish but saturate in the limit $E_{J}/E^{c2}_{J}\rightarrow\infty$. If the absolute value of these coherences in this limit is sufficiently large compared to the populations, entanglement can occur even in the regime of very strong driving.

In this section, we want to focus on this strong-driving limit to illustrate that the $\kappa$ direction of our sketch of entanglement space above is more complex than suggested so far by merely discussing the reduced Hilbert space of the two parties. To that end, we restrict the cavities to a $3\times3$ system but study the impact of modifying the relative size of the different transition matrix elements of the driving Hamiltonian.

This is not purely an academic exercise but is motivated by recent circuit-QED experiments which employ additional oscillating driving fields to essentially cut off the harmonic ladder of energy states of a stripline cavity at some level $N$~\cite{Bretheau2015}. We envision similar schemes being accomplished in our Josephson-photonics setup and thus study a so-realized $3\times3$ system while allowing for a variable $\kappa$ to change the ratio $T_{1,2}/T_{0,1}$ of the single-cavity matrix elements, see Fig.~\ref{fig_Fig_3}.

The dashed lines indicate here some special realizations of $3\times3$ systems: (i) an otherwise harmonic system ($\kappa\rightarrow0$) cut off to $3\times3$, (ii) a symmetric $3\times3$ system ($\kappa=2-\sqrt{2}$), (iii) the ``native'' Josephson-photonics $3\times3$ system ($\kappa=3-\sqrt{3}$), and (iv) the $2\times2$ system ($\kappa=2$) also natively realized in Josephson photonics without additional fields.
The results depicted in Fig.~\ref{fig_Fig_3} show that whether $\rho^{\mathrm{st}}$ in the strong-driving limit is entangled or not strongly depends on the value of $\kappa$ and thus on the ratio of the matrix elements. Among the indicated special cases, only the native Josephson-photonics $3\times3$ system is entangled in the strong-driving limit.

Actually shown in Fig.~\ref{fig_Fig_3} is the logarithmic negativity
expressed in terms of the density matrix elements as  $E_{N}[\rho^{\mathrm{st}}]=\mathrm{max}\lbrace0,\log_{2}[1+2(|\rho^{\mathrm{st}}_{00,22}|-\rho^{\mathrm{st}}_{02,02})\rbrace$. Here, $E_{N}[\rho^{\mathrm{st}}]=0$ is both necessary and sufficient for separability: if $E_{N}[\rho^{\mathrm{st}}]$ vanishes, $\rho^{\mathrm{st}}$ can be written as a sum of a separable (diagonal) state and a $2\times2$ state formed by the levels $\ket{0}$ and $\ket{2}$ which turns out to be separable if $|\rho_{00,22}|\le\rho_{02,02}$.

\subsection{Entanglement dynamics}
\label{subsec_Entanglement_dynamics}

The time-resolved statistics of photon emission events from the cavities has recently been studied experimentally and theoretically for Josephson-photonics setups~\cite{Grimm2015,Parlavecchio2015,Parlavecchio2016,Trif2015,Dambach2015,Dambach2016,Kubala2016,Leppaekangas2015,Leppaekangas2016}. Borrowing tools from quantum optics~\cite{Walls1994,Molmer1993,Plenio1998}, in particular the second-order correlation function $g^{(2)}(\tau)$,  provides a deep insight into the strongly correlated quantum dynamics of the oscillator subsystems.
Following this work, it seems quite natural to also consider the entanglement dynamics after a single photon emission event is observed. The photon emission process from oscillator $a$ is described by a jump operator $\mathfrak{J}_{a}$ acting on the density operator, $\mathfrak{J}_{a}=\gamma_{a}a\rho a^{\dagger}$.

The dynamics of entanglement is also often studied in the context of decoherence. Then, an initially prepared entangled state interacts with an environment leading to a decay and (oftentimes) a sudden death of entanglement after a finite time~\cite{Jakobczyk2002,Yu2004,Jakobczyk2004,Yu2009}.

\begin{figure}
\centering
\includegraphics[width=1.0\columnwidth]{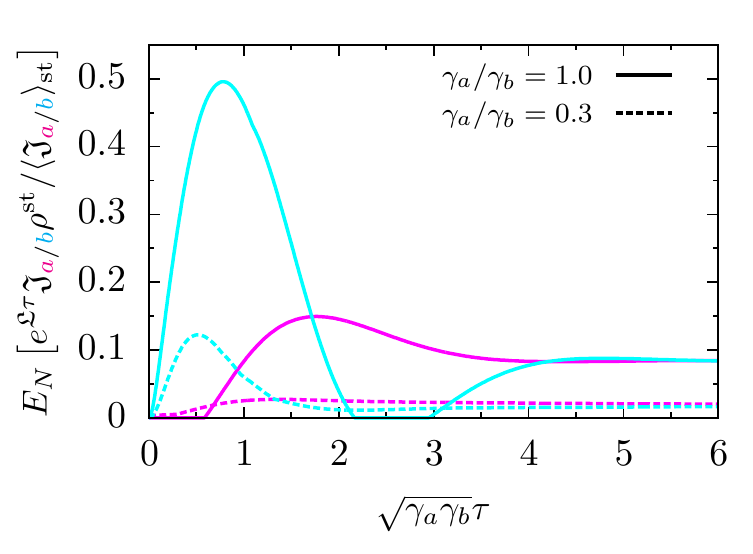}
\vspace{-0.45cm}
\caption{
Dynamics after a photon measurement of the logarithmic negativity $E_{N}$ for a $16\times2$ system initially in steady state for $E_{J}/E^{c2}_{J}=1.75$. Observing a photon leaving oscillator $a$ (magenta) or $b$ (cyan) at $\tau=0^{-}$ leads to a complete loss of entanglement. Features like entanglement sudden death and revival observable in the subsequent dynamics of $E_{N}$ (for symmetric decay) can be traced back to the presence of the two-level system and vanish for strong asymmetry ($\gamma_{a}/\gamma_{b}=0.3$).
}
\label{fig_Fig_4}
\end{figure}

Here, we consider again the special case of a $16\times2$ system moderately driven, $E_{J}/E^{c2}_{J}=1.75$, below the entanglement-separability threshold to its steady state. Then at $\tau=0^{-}$, a photon leaving the system is detected. Figure~\ref{fig_Fig_4} pictures the subsequent time evolution of the logarithmic negativity after emission  from cavity $a$ or $b$ for $\gamma_{a}/\gamma_{b}=1$ (solid lines) and $\gamma_{a}/\gamma_{b}=0.3$ (dashed lines). In all these cases, the two oscillator subsystems are separable at $\tau=0^{+}$ immediately after the photon emission event. Note that $E_{N}=0$ here is sufficient for separability after a photon measurement, which can be proven by a similar reasoning as outlined above in Sec.~\ref{subsec_Restricted_Hilbert} for the steady state of the $16\times 2$ system. If the emitted photon stems from the two-level system leaving it in its ground state, separability is obvious. Surprisingly, however, a separable state is also found if the photon is emitted from the sixteen-level system.

Let us first focus on symmetric decay rates, i.e., $\gamma_{a}/\gamma_{b}=1$. If a photon jump out of the sixteen-level system is detected, the two-level system might still be in its excited state blocking further photon creation processes and thus hindering the development of entanglement. This is why $E_{N}$ stays zero for some time after the emission event during which it becomes more and more likely that the two-level system has already relaxed before bipartite entanglement is created again. In contrast, if a photon is emitted from the two-level system, a new creation process can occur immediately, which is reflected in a strong increase in $E_{N}$ on short time scales. In this case, the entanglement features a distinct maximum but vanishes again for a finite span of time before $E_{N}$ finally approaches its steady-state value.
Note that the ``suddenness'' of death and revival features, here as well as in the decoherence dynamics of entanglement, is of course simply related to the definition of logarithmic negativity: while all density matrix elements and eigenvalues evolve smoothly, the sum of negative eigenvalues only does not.

Turning toward asymmetric decay rates, $\gamma_{a}/\gamma_{b}=0.3$, the degree of bipartite entanglement is generally lowered since the two-level system is much less populated. Furthermore, the delayed revival of entanglement is not that pronounced anymore as the probability that the two-level system is still excited after an emission event from the sixteen-level system is reduced. The fact that the vanishing of entanglement for a finite time span is lifted here completely indicates that this phenomenon can be traced back to the harsh state-space restriction in oscillator $b$, which has here a less dramatic effect due to the large $\gamma_{b}$.

In the preceding Secs.~\ref{subsec_PPT_criterion_and_logarithmic_negativity} to~\ref{subsec_Entanglement_dynamics}, we have exploited that for the bipartite case the characterization of entanglement is comparatively simple so that even the amount of entanglement can be quantified. As a general feature, the corresponding  logarithmic negativity first increases with increasing driving to approach a maximum. For stronger driving, specific features crucially depend on the way the system is driven.
Due to the high accuracy with which these devices can be characterized and controlled, the explicit values for parameters predicted here allow now to access desired entanglement properties in future experiments,  e.g., by choosing the predicted optimal driving strength to maximize the amount of entanglement.        
   
\subsection{Map of entanglement}   

The ratio $\gamma_{a}/\gamma_{b}$ of the decay rates, as seen in the previous section, has a considerable impact on the characteristics of entanglement. Once the dimension of effective Hilbert space is determined, i.e., both the number of active oscillators and the number of accessible levels in each of these oscillators is fixed,
there still remain two parameters, $E_{J}/E^{c2}_{J}$ and $\gamma_{a}/\gamma_{b}$, which determine whether entanglement is present or not. This can be visualized in an entanglement map dividing the two-dimensional space spanned by these parameters into regions of separable or entangled states.

In the two-qubit case, used in Fig.~\hyperref[fig_Fig_5]{5(a)} as a simple example, the PPT criterion already provides a necessary and sufficient condition for entanglement. Immediately, the corresponding steady-state entanglement map in Fig.~\hyperref[fig_Fig_5]{5(a)} results, with entangled states (NPT) below and separable states (PPT) above $E_{J}/E^{c2}_{J}= (\gamma_{a}^2+\gamma_{b}^2)/(\gamma_{a}\gamma_{b})$. The phenomenon of such an entanglement-separability threshold has also been observed in a number of similar steady-state systems, see, e.g., Refs.~\cite{Hartmann2006,Huelga2007,Lambert2007,ContrerasPulido2008,Li2009}.

Determining the exact boundary line between entangled and separable states for an arbitrary bipartite system is usually more difficult due to the lack of criteria which are both necessary and sufficient for the detection of entanglement. Nonetheless, the basic structure of such an entanglement map in the bipartite case is always quite trivial as any state falls in either of only two classes. This, however, changes drastically when turning toward multipartite systems in the following section, where the entanglement structure naturally becomes much more complex [cf. Fig.~\hyperref[fig_Fig_5]{5(b)}]. 

\section{Multipartite entanglement}
\label{sec_Multipartite_entanglement}

The structure of entanglement between more than two parties is much richer than in the bipartite case and does not represent a trivial extension of these results~\cite{Bruss2002,Terhal2002,Werner2002,Braunstein2005,Guehne2009,Horodecki2009}. Besides the basic question whether a given multipartite state is separable or entangled, we now have to specify the type of entanglement. A system consisting of $N$ parties in a nonseparable state does not necessarily have to be genuinely $N$-partite entangled but can also be $m$-separable ($1<m<N$), i.e., there exists a splitting of the $N$ parties into $m$ groupings which are separable from each other. In addition to this, even genuinely $N$-partite entangled states can further be divided into different subclasses. The classification of multipartite states, in particular multipartite \emph{mixed} states, is still a matter of intense research and, despite all efforts, it is far from being completely understood in general and is only explicitly known for very special cases.

Here, we therefore concentrate on one of these special cases, namely, a three-qubit system realized for $\kappa_q=2$ in our setup. In Sec.~\ref{subsec_Classification_of}, we briefly review a commonly used classification scheme for mixed three-qubit states before we apply this to our specific $2\times2\times2$ system in Sec.~\ref{subsec_Map_of} and chart the corresponding map of entanglement classes.

\subsection{Classification of mixed three-qubit states}
\label{subsec_Classification_of}

For pure three-qubit states, it is well known that there exist different equivalence classes of entanglement~\cite{Duer2000}, and such a classification scheme can also be extended to mixed states~\cite{Acin2001}.

A pure three-qubit state is called a \emph{fully separable state} if it can be written as $\ket{\varphi_{\mathrm{fs}}}=\ket{\psi_{a}}\otimes\ket{\psi_{b}}\otimes\ket{\psi_{c}}$. A \emph{biseparable state} is a state where one of the three parties is separable from the other two. Consequently, there exist three different classes of biseparable states depending on which of the two subsystems are grouped together: for example, $\ket{\varphi_{\mathrm{bs},a}}=\ket{\psi_{a}}\otimes\ket{\psi_{bc}}$ indicates a state where the parties $b$ and $c$ may be entangled. States which are neither fully separable nor biseparable are called \emph{genuinely tripartite entangled}. These states, however, can further be divided into two classes of inequivalent states, the so-called \emph{W-class states} $\ket{\varphi_{\mathrm{W}}}$  with representative $\ket{\mathrm{W}}=(\ket{001}+\ket{010}+\ket{100})/\sqrt{3}$ and \emph{GHZ-class states} $\ket{\varphi_{\mathrm{GHZ}}}$ represented by $\ket{\mathrm{GHZ}}=(\ket{000}+\ket{111})/\sqrt{2}$. Genuinely entangled states belonging to the GHZ class or W class cannot be transformed into one another by stochastic local operations and classical communication (SLOCC).

For mixed three-qubit states corresponding classes based on the results for pure states via convex combinations can be defined. A mixed state is considered as a \emph{fully separable state} if it can be written as a convex combination of fully separable pure states $\ket{\varphi^{j}_{\mathrm{fs}}}$, i.e., $\rho_{\mathrm{fs}}=\sum_{j}p_{j}\ket{\varphi^{j}_{\mathrm{fs}}}\bra{\varphi^{j}_{\mathrm{fs}}}$ with convex weights $p_{j}>0$ and $\sum_{j}p_{j}=1$. Accordingly, a mixed state is called \emph{biseparable} if it can be expressed as a convex sum of pure biseparable states $\ket{\varphi^{j}_{\mathrm{bs}}}$ and pure fully separable states $\ket{\varphi^{j}_{\mathrm{fs}}}$. The pure biseparable states in this sum might be entangled with respect to different partitions. Additionally, we define subclasses of biseparable mixed states where pure biseparable states in the convex sum are all separable with respect to a fixed partition, i.e., either the partition $a|bc$, $b|ac$, or $c|ab$. States which are part of the \emph{W class} can be expressed as a convex sum of pure W-class states $\ket{\varphi^{j}_{\mathrm{W}}}$, pure biseparable states $\ket{\varphi^{j}_{\mathrm{bs}}}$, and pure fully separable states $\ket{\varphi^{j}_{\mathrm{fs}}}$. All mixed states which are not covered by the classification so far are in the \emph{GHZ class}. States which belong either to the W or GHZ class are summarized as \emph{genuinely tripartite entangled states}.

\subsection{Map of entanglement classes}
\label{subsec_Map_of}

\begin{figure*}
\centering
\includegraphics[width=1.0\textwidth]{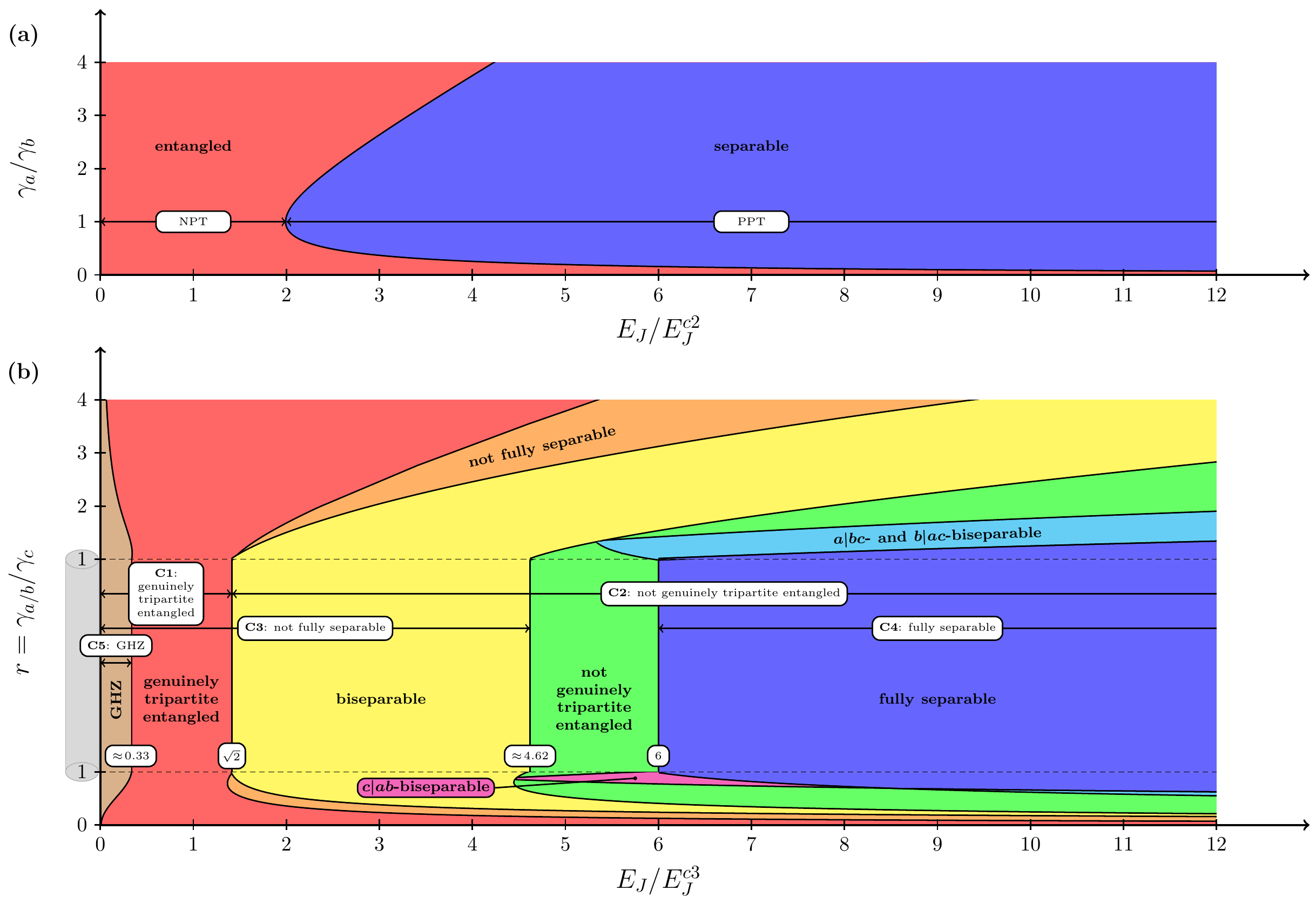}
\vspace{-0.45cm}
\caption{
(a)~Map of entanglement classes in the parameter space of the $2\times2$ system ($\kappa_{a}=\kappa_{b}=2$) in steady state spanned by the driving, $E_{J}/E^{c2}_{J}$, and the ratio of decay rates, $\gamma_{a}/\gamma_{b}$. Since the PPT criterion is both necessary and sufficient for the detection of entanglement here, any state can directly be assigned to either the class of entangled or separable states.
(b)~The corresponding map for the $2\times2\times2$ system ($\kappa_{a}=\kappa_{b}=\kappa_{c}=2$) with one differing decay rate, $\gamma_{a/b}/\gamma_{c}\neq1$. Combining the statements of different entanglement criteria (C1-C5) reveals a rich structure of different regions in parameter space which are associated with one (or several) three-qubit entanglement classes.
}
\label{fig_Fig_5}
\end{figure*}

We will now discuss how to chart a map of entanglement classes for the various stationary states of our Josephson-photonics device in the three-qubit case, $\kappa_q=2$. Figure~\hyperref[fig_Fig_5]{5(b)} shows the resulting map in the parameter space spanned by the driving strength and an additional parameter allowing for unequal damping of the cavities.  To chart this map of entanglement classes, we will not directly rely on the above introduced definitions since a straightforward rewriting of the density matrix is usually not feasible. Instead, various witnesses or other criteria which have been developed based on these definitions are employed delivering statements on the entanglement nature of the stationary state for a certain region of parameter space. These multiple statements are then eventually combined to construct the full map.

We first find that the steady-state density matrix in the three-qubit case has a simple structure,
\begin{equation}
\setlength{\arraycolsep}{-5pt}
\!\,\,\rho^{\mathrm{st}}\!=\!\!
\begin{pmatrix}
\cellcolor{gray!30}\rho^{\mathrm{st}}_{000,000}&0&0&0&0&0&0&\cellcolor{gray!30}\rho^{\mathrm{st}}_{000,111}\\
0&\cellcolor{gray!30}\rho^{\mathrm{st}}_{001,001}&0&0&0&0&\cellcolor{gray!30}0&0\\
0&0&\cellcolor{gray!30}\rho^{\mathrm{st}}_{010,010}&0&0&\cellcolor{gray!30}0&0&0\\
0&0&0&\cellcolor{gray!30}\rho^{\mathrm{st}}_{011,011}&\hspace*{10pt}$\colorbox{gray!30}{$\:\!\!\!\!\!\!_{\phantom{11}}0_{\phantom{11}}\,\:\!$}$&0&0&0\\
0&0&0&\cellcolor{gray!30}0&\cellcolor{gray!30}\rho^{\mathrm{st}}_{100,100}&0&0&0\\
0&0&\cellcolor{gray!30}0&0&0&\cellcolor{gray!30}\rho^{\mathrm{st}}_{101,101}&0&0\\
0&\cellcolor{gray!30}0&0&0&0&0&\cellcolor{gray!30}\rho^{\mathrm{st}}_{110,110}&0\\
\cellcolor{gray!30}\rho^{\mathrm{st}}_{111,000}&0&0&0&0&0&0&\cellcolor{gray!30}\rho^{\mathrm{st}}_{111,111}\\
\end{pmatrix}\!\!,\!\!\!
\label{eq_three-qubit_state}
\end{equation}
when written in a standard product basis $\lbrace\ket{000},\ket{001},\ket{010},\dots,\ket{111}\rbrace$.
Alluding to the pattern of nonzero matrix elements, states of this structure are known as X states (here, there are some additional zeros). X states have been widely studied~\cite{Yu2007,Rau2009,Ali2010,Guehne2010,Weinstein2010,Siomau2010,HashemiRafsanjani2012} with respect to entanglement and other quantum properties, in particular a subset of these states, which is called GHZ-diagonal~\cite{Duer1999}. Despite their simple structure, states of this form yield a rich pattern in the map of entanglement classes as we will see in the following.

Allowing for the decay rate of one of the qubits to differ from the other two, i.e., $r=\gamma_{a/b}/\gamma_{c}\neq1$, the density matrix elements of $\rho^{\mathrm{st}}$ in Eq.~\eqref{eq_three-qubit_state} can be explicitly found as simple functions of driving, $E_{J}/E^{c3}_{J}=E_{J}\sqrt{\kappa_{a}\kappa_{b}\kappa_{c}}e^{-(\kappa_{a}+\kappa_{b}+\kappa_{c})/2}/(\hbar\sqrt{\gamma_{a/b}\gamma_{c}})$, and asymmetry [see Eq.~\eqref{eq_analytical_results_three_qubits} in the Appendix~\hyperref[Appendix_B]{B}].

With those expressions, we can now proceed in our entanglement classification focusing first on symmetrically damped oscillators, $r=1$.\\
\textbullet~Criterion C1 is a simple witness proposed by Gühne and Seevinck~\cite{Guehne2010} based directly on the density matrix elements of a three-qubit system that allows us to detect genuinely entangled states: If $\rho$ is a biseparable state, then its matrix elements fulfill the inequality $|\rho_{000,111}|\le\sqrt{\rho_{001,001}\,\rho_{110,110}}+\sqrt{\rho_{010,010}\,\rho_{101,101}}+\sqrt{\rho_{011,011}\,\rho_{100,100}}$. Here, we find the inequality violated for  all states where $E_{J}/E^{c3}_{J}<\sqrt{2}$, implying genuine tripartite entanglement below this driving strength, see Fig.~\hyperref[fig_Fig_5]{5(b)}.\\
\textbullet~Criterion C2 based by Novo et al.~\cite{Novo2013} on a combination of PPT mixtures and permutationally invariant states, in contrast, detects states which are definitely \emph{not} genuinely entangled but biseparable at the most. A PPT mixture is defined as a convex combination of PPT states with respect to a specific partition, i.e., $\rho^{\mathrm{mix}}=p_{a}\rho^{\mathrm{mix}}_{a}+p_{b}\rho^{\mathrm{mix}}_{b}+p_{c}\rho^{\mathrm{mix}}_{c}$ with $(\rho^{\mathrm{mix}}_{j})^{T_{j}}\ge0$ $\forall j$. A permutationally invariant three-qubit state is biseparable if and only if it is a PPT mixture. Here, we can show that for $\sqrt{2}\le E_{J}/E^{c3}_{J}$ all states can be written as a convex sum of fully separable states and a permutationally invariant PPT mixture, i.e., all these states are \emph{not} genuinely tripartite entangled. For $\sqrt{2}> E_{J}/E^{c3}_{J}$, we can not make any statements.\\
\textbullet~Criterion C3 follows work by Wölk et al.~\cite{Woelk2014} to develop an entanglement witness based on the Hölder inequality which detects nonseparable states: If $\rho$ is a fully separable state, then the inequality $|\rho_{000,111}|\le\sqrt[4]{\rho_{000,000}\,\rho_{011,011}\,\rho_{101,101}\,\rho_{110,110}}$ holds. Violation is consequently a sufficient condition for $\rho$ to be \emph{not} fully separable, which indeed is the case here for all states where $E_{J}/E^{c3}_{J}<3\sqrt{\sqrt{33}-1}/\sqrt{2}\approx4.62$. This implies that $\rho^{\mathrm{st}}$ is biseparable, but not fully separable, in the regime $\sqrt{2}\le E_{J}/E^{c3}_{J}\lesssim4.62$. \\
\textbullet~Criterion C4 is one of a large number of powerful criteria that is available for states which are diagonal in the GHZ basis $\ket{\Psi^{\pm}_{j}}=(\ket{\varphi_{a}\varphi_{b}\varphi_{c}}\pm\ket{\bar{\varphi}_{a}\bar{\varphi}_{b}\bar{\varphi}_{c}})/\sqrt{2}$ with $\varphi_{l}\in\lbrace0,1\rbrace$ and $\varphi_{l}\neq\bar{\varphi}_{l}$.
 Dür et al.~\cite{Duer1999} presented a necessary and sufficient criterion to decide whether such a GHZ-diagonal state is fully separable. We first apply an invertible local transformation to $\rho^{\mathrm{st}}$ such that $\rho^{\mathrm{st}}_{000,111}=\rho^{\mathrm{st}}_{111,000}\in\mathbb{R}$, thereby not changing the separability properties. For $6\le E_{J}/E^{c3}_{J}$, it can then be shown that all these states can be written as a convex sum of obviously fully separable states and GHZ-diagonal states which turn out to be fully separable. Again, we can make no statement for $6> E_{J}/E^{c3}_{J}$. For states in the regime $4.62\lesssim E_{J}/E^{c3}_{J}<6$, we can thus only state that they are not genuinely tripartite entangled.\\
\textbullet~Criterion C5 allows the detection of GHZ-class states and was derived by Eltschka and Siewert~\cite{Eltschka2013} on the basis of GHZ-symmetric states~\cite{Eltschka2012}: If $\langle W\rangle<0$, where  $W=(3/4)\mathbb{1}-3/(\nu_{0}^{2}-2\nu_{0}+4)\pi_{\mathrm{GHZ}_{+}}-3/(\nu_{0}^{2}+2\nu_{0}+4)\pi_{\mathrm{GHZ}_{-}}$ with $\pi_{\mathrm{GHZ}_{\pm}}=(\ket{000}\pm\ket{111})(\bra{000}\pm\bra{111})/2$ and $-1\le\nu_{0}\le1$, then $\rho$ is part of the GHZ class. This criterion yields that $\rho^{\mathrm{st}}$ is a GHZ-class state in the regime $E_{J}/E^{c3}_{J}\lesssim0.33$.

Dropping the restriction to equally damped oscillators, for $r\neq1$ the structure of entanglement becomes slightly modified as new regimes occur. Firstly, a gap arises between the regime boundaries based on C1 and C2 so that
a region develops where states are known (by C3) to be \emph{not} fully separable, but the type of entanglement could not be determined.
Secondly, we can identify two additional regimes of biseparable states where the states are separable with respect to a fixed partition. This is shown again by rewriting $\rho^{\mathrm{st}}$ as a sum of GHZ-diagonal states and fully separable states. Building on work of Dür et al.~\cite{Duer1999}, we can prove that these GHZ-diagonal states are biseparable with respect to the corresponding partition. As apparent in Fig.~\hyperref[fig_Fig_5]{5(b)}, the overlap of these two regimes eventually forms the regime of fully separable states.

Intuitively one might think that a strong symmetry between the three oscillators would favor genuine tripartite entanglement. The results in Fig.~\hyperref[fig_Fig_5]{5(b)} reveal, however, that in the near-symmetric case, $r\approx1$, the transition from genuine tripartite entanglement to biseparability and eventually further to full separability already occurs at comparatively moderate driving. Tuning the Josephson energy $E_{J}$ in this near-symmetric case allows one to access a three-qubit state with the desired entanglement properties. For increasing asymmetry, the boundary lines between these regimes are shifted towards higher values of $E_{J}$ and we end up in an entangled state even for very strong driving. The blockade of further excitation and de-excitation processes once a photon has left the system from triple occupation $\ket{111}$ is particularly effective for strong asymmetry and protects the genuine tripartite entanglement against the impact of strong driving.
The influence of the driving strength on the entanglement properties can again be understood by considering the relative size of the coherences $|\rho^{\mathrm{st}}_{000,111}|=|\rho^{\mathrm{st}}_{111,000}|$ and the populations $\rho^{\mathrm{st}}_{j,j}$ (cf. explicit discussion for the $2\times2$ realization in Sec.~\ref{subsec_Restricted_Hilbert}).
In the weak-driving limit, the coherences are dominant compared to the populations implying a strong quantum character in form of tripartite entanglement. Conversely, the coherences become small for strong driving, while the populations saturate leading to a more classical behavior which is reflected in full separability.

One important question which has not been addressed so far is whether there are also genuinely tripartite entangled states which belong to the W class. This has not yet been excluded by the criteria discussed here for the red region in Fig.~\hyperref[fig_Fig_5]{5(b)}. Possibilities to show that the genuinely tripartite entangled states belong to the W class are unfortunately scarcely available. In particular, W-class states cannot be detected by witnesses in general as these are not designed to prove that a state lies inside a convex set~\cite{Guehne2009}.

To summarize this part, we have analyzed the surprisingly rich entanglement map of the photonic states which occur in form of stationary states in driven multimode cavity-Josephson circuits. This is even more striking considering the simple structure of these states [Eq.~\eqref{eq_three-qubit_state}]. The fact that this map is spanned by only two experimentally easily accessible parameters underlines the potential of these circuits as well-controlled platforms for entanglement generation beyond bipartite situations. In particular, their operation does not require any complicated pulse shaping of external microwaves. However, this map still contains unexplored territory which may trigger further theoretical research to develop corresponding entanglement criteria.

\section{Experimental situation}
\label{sec_Experimental_situation}

Current experiments in Josephson photonics have already realized two-cavity setups and taken first steps in tackling the nonclassicality of microwave emission~\cite{Parlavecchio2015,Parlavecchio2016}. We now shortly turn to a discussion to what extent the promised versatility of Fig.~\hyperref[fig_Fig_1]{1(c)} can actually be fulfilled and what improvements and modifications may still be necessary.

Constructing setups of three or more cavities and tuning the bias voltage $V$ between the corresponding resonances to switch directly from a bipartite to a multipartite system does not seem to pose a principle challenge. Besides the number of active parties, also the Josephson energy $E_{J}$, i.e., the class of entanglement, is easily tunable over a few orders of magnitude via the magnetic flux when using a SQUID configuration for the Josephson junction.

By contrast, the dimension of Hilbert space determined by the parameters $\kappa_{q}$ is currently fixed by design.
While earlier experimental realizations were limited to the low-impedance regime $\kappa_{q}\ll1$, recent progress already makes it possible to reach values up to $\kappa_{q}\approx1.6$~\cite{Portier2016}. Achieving a $\kappa_{q}$ value which exactly matches one of the special values restricting the Hilbert space to a finite number of levels is actually not that crucial. Assuming  $N \times M$ systems above is convenient for calculations and interpretation. However, except for extremely strong driving there will hardly be a difference between a near suppression and an exact vanishing of a certain transition matrix element [cf. Fig.~\hyperref[fig_Fig_2]{2(a)} and also the discussion in Ref.~\cite{Souquet2016}, showing that  deviations in the populations in a $2\times2$ system are of order ${\cal{O}}(\delta \kappa^2)$]. Nonetheless, if tunability is desired, employing SQUIDs at the end of a cavity or meta-material striplines constructed from long SQUID arrays~\cite{Jung2014}, the effective cavity length (and therefore the resonance frequency) can be changed by magnetic fluxes. Moreover, the actual  defining equation $\kappa_{q}=E_C/(\hbar\omega_{q}^{(0)})$, where $E_{C}=2e^{2}/C_{q}$ is the charging energy, corresponds to  $\kappa_{q} =(2e^{2}/\hbar)\sqrt{L_{q}/C_{q}}$ for the fundamental $\lambda/4$ mode $\omega^{(0)}_{q}$. Accessing higher modes instead, therefore also gives immediate in-situ access to lower $\kappa_q$ values within currently existing devices.

\begin{figure}
\centering
\includegraphics[width=1.0\columnwidth]{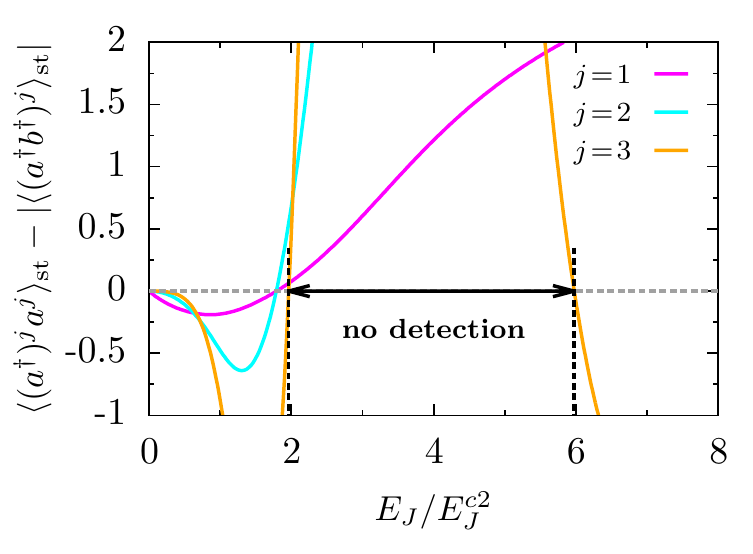}
\vspace{-0.45cm}
\caption{
Witnesses $\langle W_{j}\rangle_{\mathrm{st}}=\langle(a^{\dagger})^{j}a^{j}\rangle_{\mathrm{st}}-|\langle(a^{\dagger}b^{\dagger})^{j}\rangle_{\mathrm{st}}|$ for $j=1$, $2$, and $3$ detecting steady-state entanglement in the symmetric $16\times16$ system ($\kappa_{a}=\kappa_{b}\approx0.23$) for values of $E_{J}/E^{c2}_{J}$  where $\langle W_{j}\rangle_{\mathrm{st}}<0$. All three witnesses fail in detecting nonseparability in a regime of moderate driving between $E_{J}/E^{c2}_{J}\approx1.96$ and $5.97$, where the presence of entanglement has already been proven by the PPT criterion (cf. Fig~\ref{fig_Fig_2}).
}
\label{fig_Fig_7}
\end{figure}

Our theoretical investigations have been restricted to a small portion of the "space of entanglement phenomena" [Fig.~\hyperref[fig_Fig_1]{1(c)}], namely, the full bipartite case and the low-dimensional $2\times2\times2$ Hilbert space in the tripartite case, mainly due to the fact that only here a classification scheme is known and implementable criteria to detect these classes are available. The criteria used in our studies of bipartite and multipartite entanglement in the previous two sections, moreover, all require the full knowledge of the density matrix $\rho$. In an experiment, $\rho$ can be reconstructed on the basis of quantum state tomography~\cite{Leonhardt1997}, where the density matrix of an unknown state is fully determined by repeatedly performing measurements in different bases on an ensemble of identical copies of this state. In principle, investigating steady-state entanglement, here, such identical copies come for free. However,
the whole procedure is challenging and, moreover, costly in terms of time, which as will be discussed below will pose severe limitations on implementing it in our setup.
Particularly for systems with large Hilbert spaces, the knowledge of a quantum state $\rho$ in an experimental situation will therefore usually be incomplete as based on a small number of observables only. For that reason, a large number of entanglement criteria in terms of directly measurable observables have been proposed. Sometimes these are simple witnesses relying on a single inequality only (see, e.g., Refs.~\cite{Terhal2000,Bourennane2004,Korbicz2005,Hillery2006,Huber2010,Woelk2014,Guehne2010}), but also more elaborate detection schemes are known which, e.g., relate the PPT criterion of a given state $\rho$ to the positivity of a corresponding (infinite) matrix of moments~\cite{Shchukin2005,Miranowicz2009}. Whether one of these criteria is successful in detecting entanglement or not strongly depends on the structure of $\rho$. In principle, however, there exists an entanglement witness for any entangled state $\rho$~\cite{Horodecki1996}.
To illustrate the limitations of simple witnesses in detecting entanglement, we apply in Fig.~\ref{fig_Fig_7} the first three witnesses, $j\in\lbrace1,2,3\rbrace$, of the family $\langle W_{j}\rangle_{\mathrm{st}}=\langle(a^{\dagger})^{j}a^{j}\rangle_{\mathrm{st}}-|\langle(a^{\dagger}b^{\dagger})^{j}\rangle_{\mathrm{st}}|$ [cf. Eq.~\eqref{eq_witness_bipartite} for $j=1$]~\cite{Hillery2006} to the steady state $\rho^{\mathrm{st}}$ of the symmetric $16\times16$ system. They detect entanglement, $\langle W_{j}\rangle_{\mathrm{st}}<0$, in the weak-driving and strong-driving regime only, however, not in an intermediate regime between $E_{J}/E^{c2}_{J}\approx1.96$ and $5.97$, where entanglement indeed is present (cf. Fig.~\ref{fig_Fig_2}).

\begin{figure*}
\centering
\includegraphics[width=1.0\textwidth]{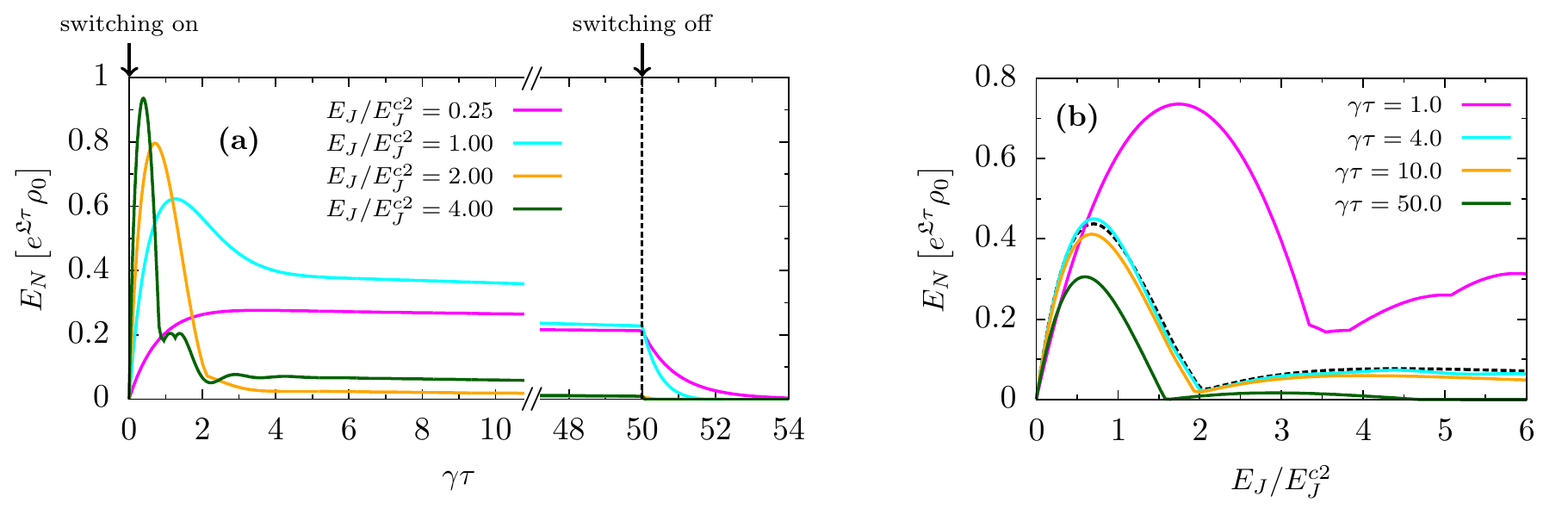}
\vspace{-0.45cm}
\caption{
The steady-state values of entanglement of Sec.~\ref{sec_Bipartite_entanglement} are observable despite the presence of weak voltage noise, $\gamma_{J}/\gamma=0.01$. Data are shown for a symmetric $3\times3$ system ($\kappa_{a}=\kappa_{b}=3-\sqrt{3}$, $\gamma_{a}=\gamma_{b}=\gamma$). (a)~On a time scale of a few $1/\gamma$ after switching on the device, the system approaches a quasistationary state with  logarithmic negativity $E_{N}$ close to the steady-state results without voltage noise [cf. (b)]. $E_{N}$ then slowly decays exponentially with a rate $\propto\gamma_{J}$. Switching off driving again at $\gamma\tau=50$ leads to a fast loss of the residual entanglement due to photon leakage from the cavity. (b)~The corresponding logarithmic negativity now as a function of $E_{J}$ at different times after switch-on in comparison to the steady-state results for vanishing voltage noise ($\gamma_{J}=0$, dashed line).
}
\label{fig_Fig_6}
\end{figure*}

Leakage of excited photons from the resonators into the electromagnetic environment is the dominant but not the only source of decoherence in an actual experimental realization. As discussed in Ref.~\cite{Gramich2013}, local voltage fluctuations at the Josephson junction associated with a rate $\gamma_{J}$ are comparatively weak, $\gamma_{J}/\gamma_{q}\ll1$, in experiments~\cite{Hofheinz2011,Chen2014}. While they can be safely disregarded for some observables~\cite{Gramich2013,Armour2015,Souquet2016}, they have a non-negligible impact on the entanglement properties in steady state. Describing the effect of voltage noise requires an extended model~\cite{Gramich2013,Armour2015} with an extra degree of freedom for the number of Cooper pairs $N$ that have passed the tunnel element. The bracketed term in the Hamiltonian [Eq.~\eqref{eq_Hamiltonian_RWA}] is here replaced by $(\prod_{q}e^{i\eta}a_{q}^{\dagger}+\prod_{q}e^{-i\eta}a_{q})$ with $e^{i\eta}=\sum_{N}\ket{N}\bra{N+1}$, where the phase difference $\eta$ across the junction and $N$ form a pair of conjugated variables, $[\eta,N]=i$. The quantum master equation [Eq.~\eqref{eq_quantum_master_equation}] is then modified by the additional dissipator $\gamma_{J}(2N\rho N-N^{2}\rho-\rho N^{2})/2$, including the decohering effects of voltage fluctuations on the system's dynamics. To analyze how voltage noise affects the entanglement properties in steady state, we study in Fig.~\ref{fig_Fig_6} the logarithmic negativity $E_{N}$ for a $3\times3$ system and symmetric decay rates ($\gamma_{1}=\gamma_{2}=\gamma$) with $\gamma_{J}/\gamma=0.01$ during a switching-on/off process of the Josephson-photonics device.
We assume for $\rho(\tau=0)=\rho_{0}$ that the two cavities are initially in their ground state and there is a well-defined phase difference across the Josephson junction, $\langle (e^{\pm i\eta})^{j}\rangle=1$ with $j\in\mathbb{N}$. Figure~\hyperref[fig_Fig_6]{7(a)} pictures $E_{N}$ as a function of time for different values of the driving strength $E_{J}$ after switching on the device at $\tau=0$ by instantaneously turning up the driving from $E_{J}/E^{c2}_{J}=0$ to the respective value.
In practice, the switching process can be performed considerably quicker than other relevant timescales, in particular $1/\gamma$~\cite{Leppaekangas2015}.

The logarithmic negativity increases immediately and approaches, on a time scale of a few $1/\gamma$, a quasistationary state decaying exponentially but only very slowly with rates of the order $\gamma_{J}$. At $\gamma\tau=50$, we finally switch of the device again by instantaneously reducing the driving to zero, resulting in a fast vanishing of the entanglement on time scales associated with $1/\gamma$. In Fig.~\hyperref[fig_Fig_6]{7(b)}, $E_{N}$ is additionally plotted as a function of the driving $E_{J}/E^{c2}_{J}$ at different moments in time after switching on the device. At $\gamma\tau\approx4.0$, the system has already reached its quasistationary state and the corresponding logarithmic negativity almost coincides with the reference values for vanishing $\gamma_{J}$, where the system is in a true stationary state (dashed line, cf. Fig.~\ref{fig_Fig_2}). Due to the impact of voltage noise, the degree of entanglement is henceforth continuously reduced and after a long, but finite, waiting time the system would eventually end up in a completely disentangled state. Nonetheless, we can conclude that for weak voltage noise there exists a time span sufficiently long for experimental observation where the multifaceted entanglement properties discussed above for vanishing $\gamma_{J}$ are definitely present before switching off and restarting the device becomes necessary.

\section{Conclusions and outlook}
\label{sec_Conclusions_and}

To study the complexity of entanglement phenomena in multipartite systems is a challenging and in many respects a still-unsolved problem, both in terms of theoretical concepts and in terms of experimental realizations.
In this paper, we demonstrated that Josephson-photonics devices which combine the basic elements of circuit QED in form of a voltage-biased Josephson junction interacting with an array of microwave cavities may serve as simple and versatile sources for the creation of entangled photons in the microwave up to the low terahertz regime. The way these photonic states are created is strikingly simple and does not require complicated pulse shaping. The relevant parameters of these devices are well-characterized and their operation is controllable to such a degree that a small set of experimental knobs allows one to access a broad variety of stationary photon states with specific entanglement properties.

In this work, we provided quantitative predictions for the amount of entanglement for the bipartite case  which are of direct relevance for experiments currently being implemented. More generally, we showed how to design the number of entangled parties and the structure of their respective Hilbert spaces by varying experimentally accessible parameters. This way, various classes of entangled multipartite photon states can be realized. Specific attention has been then spent on multipartite entanglement of two-state systems since for this specific case advanced entanglement criteria have been developed recently. However, it turned out that even for these cases our theoretical understanding is far from complete. The type of driven, dissipative mesoscopic circuits considered here may thus trigger further research to better understand entanglement in more complex situations.

\emph{For future experiments}, one important question will be the best choice of witness. Complexity or ease of measurement will have to be balanced with the power of various possible witnesses to distinguish between different classes and detect entanglement in a wider or more restricted range.
This links to the challenge of mitigating the debilitating impact of low-frequency noise on entanglement; either by choosing witnesses which can be measured during the time of quasistationarity, by further improving phase stability possibly via phase-locking schemes, or by exploiting observables and measurement schemes insensitive to phase noise (cf. Franson-interferometric schemes~\cite{Franson1989,Leppaekangas2016b}).

In that context, we want to emphasize that in the current work we concentrated on questions concerning the entanglement of cavity modes. Various closely related questions can be studied in the frequency- and/or time-domain directly for output modes, including detailed modeling of filtering and photon detection.
Related to these issues are possible limitations of our results inherent to the modeling of dissipation by a standard quantum-optical Lindbladian. How  and when to improve on this modeling is an important topic for further studies.

\emph{For applications} as an entanglement source, an important feature distinguishing Josephson-photonics devices from other circuit-QED setups is the continuous mode of operation, which suggests particular potential as  high-intensity source. Operating several Josephson junctions in a parallel, self-synchronized manner may conceivable serve this purpose.

\emph{For abstract entanglement theory}, Josephson photonics constitute one powerful example that even simple
Hamiltonians can dynamically generate a wealth of complex entanglement phenomena, worthy of interest.
This may trigger efforts to characterize and understand the entanglement properties of states which are naturally occurring (as steady states) of similar dynamical systems as studied here; thereby complementing those where the investigated states are chosen arbitrarily or according to other criteria.
Particularly fascinating is the question to what extent it is possible to directly connect the structure of the generating Hamiltonian to the entanglement class of the eventually resulting mixed states. For instance, instead of a tripartite Hamiltonian of the $a^\dagger b^\dagger c^\dagger + \mathrm{c.c} $ type as at the $2eV = \hbar (\omega_a + \omega_b + \omega_c)$ resonance of Josephson-photonics, we may consider a $a^\dagger b^\dagger + b^\dagger c^\dagger + a^\dagger c^\dagger + \mathrm{c.c} $ Hamiltonian, which is not realizable in pure form in the current setup. [There would be competing $(a^\dagger)^2$ terms among other corrections.] We may then ask whether this can be related to entanglement of GHZ or W type, respectively, and try to pose and answer similar questions for other multipartite systems.

\begin{acknowledgments}
The authors thank A.~D. Armour, J. Leppäkangas, F. Portier, and S. Wölk for valuable discussions.
This work was supported by the Deutsche Forschungsgemeinschaft (DFG) through Grant No. AN336/6-1 and SFB/TRR21
as well as by the Center for Integrated Quantum Science and Technology (\smash{$\mathrm{IQ}^{\mathrm{ST}}$}). S.D. acknowledges financial support from the Carl-Zeiss-Stiftung.
\end{acknowledgments}

\section*{Appendix A: Excitations in a subset of oscillators}
\label{Appendix_A}

Selecting the bias condition $\omega_{J}=2eV/\hbar=\omega_{a}+\omega_{b}$ in a setup of three cavities (denoted $a$, $b$, and $c$), only the oscillators $a$ and $b$ are directly involved in the fundamental creation/absorption process initiated by a tunneling Cooper pair. The presence of the passive oscillator $c$, however, results in a renormalization of this process in terms of the Bessel function $J_{0}$ in the RWA Hamiltonian close to that resonance (in a frame rotating with $\omega_{J}$):
\begin{equation}
H\!=\!\frac{\tilde{E}_{J}}{2}\!:\!\!\!\;\left(a^{\dagger}b^{\dagger}\!+\!ab\right)\!\frac{J_{1}\!\!\left(\!\sqrt{\!4\kappa_{a}n_{a}}\right)\!\!\!\;J_{1}\!\!\left(\!\sqrt{\!4\kappa_{b}n_{b}}\right)\!\!\!\;J_{0}\!\!\!\;\left(\!\sqrt{\!4\kappa_{c}n_{c}}\right)}{\sqrt{\!\kappa_{a}n_{a}}\sqrt{\!\kappa_{b}n_{b}}}\!:.
\tag{A1}\label{eq_Hamiltonian_nonactive_cavities}
\end{equation}
Here, $a^{(\dagger)}$, $b^{(\dagger)}$, and $c^{(\dagger)}$ are the corresponding creation (annihilation) operators and $\tilde{E}_{J}=E_{J}\sqrt{\kappa_{a}\kappa_{b}}e^{-(\kappa_{a}+\kappa_{b}+\kappa_{c})/2}$. In the zero-temperature limit, oscillator $c$ is in its ground state and consequently backaction on the fundamental creation/annihilation process does not occur, i.e., $J_{0}(\sqrt{4\kappa_{c}n_{c}})\rightarrow1$. The Hamiltonian of a three-cavity system with passive oscillator $c$ then equals the Hamiltonian of a two-cavity setup [cf. Eq.~\eqref{eq_Hamiltonian_RWA} for $q=a$, $b$] except for a renormalized $E_{J}$.

The structure of the Hamiltonian in Eqs.~\eqref{eq_Hamiltonian_nonactive_cavities} and \eqref{eq_Hamiltonian_RWA} relies on the existence of a single resonant process. If some of the modes are degenerate, there will be different competing resonant processes and the RWA Hamiltonian will be a sum of several terms of a form similar to that in Eqs.~\eqref{eq_Hamiltonian_nonactive_cavities} and \eqref{eq_Hamiltonian_RWA}.

\section*{Appendix B: Two-/three-qubit system in steady state}
\label{Appendix_B}

Analytical results for the density matrix $\rho$ can be found in the $2\times2$ and $2\times2\times2$ Hilbert space by explicitly solving the linear system of first-order differential equations, $\mathrm{d}\rho/\mathrm{d\tau}=\mathfrak{L}\rho$. In steady state, i.e., $0=\mathfrak{L}\rho^{\mathrm{st}}$, the nonzero matrix elements for the $2\times2$ system written in the product basis $\lbrace\ket{00},\ket{01},\ket{10},\ket{11}\rbrace$ read
\begin{equation}
\begin{aligned}
\rho^{\mathrm{st}}_{00,00}&=[1 + (2 + d_{2}^2) r_{2}^2 + r_{2}^4]/\mu_{2},\\
\rho^{\mathrm{st}}_{01,01}&=d_{2}^2 r_{2}^4/\mu_{2},\\
\rho^{\mathrm{st}}_{10,10}&=d_{2}^2/\mu_{2},\\
\rho^{\mathrm{st}}_{11,11}&=d_{2}^2 r_{2}^2/\mu_{2},\\
\rho^{\mathrm{st}}_{00,11}&=\left(\rho^{\mathrm{st}}_{11,00}\right)^{\!*}\!=id_{2} r_{2}(1 + r_{2}^2)/\mu_{2},
\end{aligned}
\tag{B1}\label{eq_analytical_results_two_qubits}
\end{equation}
where $d_{2}=E_{J}/E^{c2}_{J}$, $r_{2}=\gamma_{a}/\gamma_{b}$, and $\mu_{2}=(1 + d_{2}^2) (1 + r_{2}^2)^2$.

For the $2\times2\times2$ system allowing for one decay rate to differ from the other two, i.e., $\gamma_{a/b}\neq\gamma_{c}$, the corresponding nonzero matrix elements in the product basis \pagebreak$\lbrace\ket{000},\ket{001},\ket{010},\dots,\ket{111}\rbrace$ are given by
\begin{equation}
\begin{aligned}
\rho^{\mathrm{st}}_{000,000}&=2 (1 + r_{3}^2) [1 + (4 + d_{3}^2) r_{3}^2 + 4 r_{3}^4],/\mu_{3}\\
\rho^{\mathrm{st}}_{001,001}&=4 d_{3}^2 r_{3}^6/\mu_{3},\\
\rho^{\mathrm{st}}_{010,010}&=\rho^{\mathrm{st}}_{100,100}=d_{3}^2 (1 + 3 r_{3}^2)/\mu_{3},\\
\rho^{\mathrm{st}}_{011,011}&=\rho^{\mathrm{st}}_{101,101}=2 d_{3}^2 r_{3}^4/\mu_{3},\\
\rho^{\mathrm{st}}_{110,110}&=d_{3}^2 (1 + r_{3}^2)/\mu_{3},\\
\rho^{\mathrm{st}}_{111,111}&=2 d_{3}^2 r_{3}^2 (1 + r_{3}^2)/\mu_{3},\\
\rho^{\mathrm{st}}_{000,111}&=\left(\rho^{\mathrm{st}}_{111,000}\right)^{\!*}\!=i2 d_{3} r_{3} (1 + r_{3}^2) (1 + 2 r_{3}^2)/\mu_{3},
\end{aligned}
\!\!\!\!
\tag{B2}\label{eq_analytical_results_three_qubits}
\end{equation}
\pagebreak
where $d_{3}=E_{J}/E^{c3}_{J}$, $r_{3}=\gamma_{a/b}/\gamma_{c}$, and $\mu_{3}=2 (1 + r_{3}^2) (1 + 2 r_{3}^2)^2 + d_{3}^2 (3 + 11 r_{3}^2 + 8 r_{3}^4 + 4 r_{3}^6)$.\vspace*{2.94cm}

\bibliography{references}

\end{document}